\documentstyle[epsf]{l-aa}
\begin{document}
\thesaurus{09(19.82.1)}
\title{On Velocity and Intensity Line Asymmetries} 
\author{M. Gabriel}
\offprints{M.G}
\institute{Institut d'Astrophysique de l'Universit\'e de Li\`ege, 
5, avenue de Cointe, B-4000 Li\`ege, Belgium.}
\date{Received  ,Accepted  }
\maketitle 
\begin{abstract}
We show that, if solar 5 min. oscillations are excited by convection in the 
upper layers of the convective envelope, it is impossible to explain the 
opposite line asymmetries observed in the velocity and intensity spectra with 
assumptions on the dissipations which reduce the problem to a second order one.
The interpretation of that observation requires to solve the full 
non-adiabatic problem which is of the fourth or sixth order. We also analyze 
the causes of line asymmetries in the frame of the general problem and we show 
that to locate the source, it is better to study line asymmetries not too far 
from line centers.
 \end{abstract}
\section{Introduction}
Duvall et al. (1993) have discovered that lines of the solar acoustic 
power spectrum show asymmetries which have opposite signs in the velocity 
and the intensity signals. Following Abrams and Kumar (1996), a line will be 
said to have a positive (negative) asymmetry if it has more (less) power  on 
the high frequency-side than on the low-frequency one. The velocity lines show 
a negative asymmetry while the intensity ones show a positive one though the 
amount of asymmetry varies with frequency. This finding has been recently 
confirmed by  MDI which is one of the SOHO experiments. 

The line asymmetry was also predicted theoretically (Gabriel 1992, 1993, 1995)
and discussed by Kumar (1994), Lou and Fan (1995), Roxburgh and Vorontsov 
(1995), Abrams and Kumar (1996), Rast and Bogdan (1997a) and Nigam et al. 
(1997) but in nearly all cases for the velocity signal only.

It is considered that solar p-modes are non-linearly excited by convection in 
the upper layers of the convective envelope. The theory  was 
originally proposed by Goldreich and Keeley (1987) and further developed by 
Goldreich and Kumar (1988, 1990) and Goldreich et al. (1994) (see also Osaki 
(1990) and Musielak (1994)). Recently Rast (1997b, 1997c) has proposed a 
slightly different mechanism in which excitations are  associated with new 
down-flow plume formation. However,so far, these theories have been unable to 
predict the location and the width of the excitation zone which must be found 
from the interpretation of observations. Attempts have been made by Kumar 
(1994) and Abrams and Kumar (1996). 

Until recently, no attempt had been made to use intensity line asymmetries. 
The reason for this situation is easy to understand. Velocity measurements are 
much easily connected to theory as even the simple adiabatic theory makes 
predictions concerning the velocity. In the studies of line asymmetries done 
so far (with the exception of Kumar 1994), dissipations have however been 
taken into account but in a very rough 
way which allows, with the Cowling approximation, to keep a second order 
boundary value problem which is much easier to handle than the 
full  fourth order non-adiabatic one. Even if we can question 
the validity of the approximations done to keep a second order problem when it 
comes to make theoretical predictions accurate enough for the interpretation of
 observations, such predictions can be done and it is tempting to confront 
them with observations. However the perturbation of the luminosity does not 
appear in those simplified problems. Moreover the observations are often made 
in a frequency interval and it is not obvious whether the intensity 
fluctuations may be simply associated to the perturbations of 
the total luminosity obtained by theoretical computations. Even if this is 
allowed, we have to solve a fourth order problem the theoretical basis of 
which are poorly understood as it requires to solve the linear interactions 
between pulsation and convection. It is probably for this reason that attempts 
 trying to explain the opposite line asymmetries in velocity and intensity 
spectra with second order boundary value problems, have been made  only 
recently  after the confirmation of the discovery by MDI. To do that, it is 
necessary to introduce one more hypothesis which links
 the intensity fluctuations to either the Lagrangian or the Eulerian 
temperature (or pressure) perturbations.

In this note we prove that the solution of second order problems leads always  
the same line asymmetries for the two variables. It is therefore
  impossible to explain both the velocity and intensity observations with such 
simplified problems. Their interpretation requires to solve the full 
non-adiabatic equations.
\section{Equations of the problem}
To study  the problem of pulsations excited by convection, we have to solve a 
non-homogeneous linear problem (Gabriel 1993):
\begin{equation}
\frac{d \vec{Y}}{dr} = A \vec{Y} + \vec{F}
\end{equation}
The homogeneous system is given by the equations of non-adiabatic stellar 
stability 
and $\vec{F}$ is the forcing term produced by convection. $A$ is a matrix, 
function 
of the distance to the center and of the oscillation frequency which indeed 
takes real values only (while the eigenvalues of the homogeneous problem are 
complex). The solutions of this system have to fulfill the same boundary 
conditions as the homogeneous one. For radial oscillations and in the Cowling 
approximation for non-radial ones, the system is of dimension four, while when 
the perturbation of the potential is taken into account in the non-radial case 
the problem is of the sixth order. 

The solution of that system is given by (Gabriel 1993)
\begin{equation}
\vec{Y(r)} = \int_0^R G(r,r'). \vec{F}(r') dr'
\end{equation}
where $G(r,r')$ is the Green function matrix of the homogeneous system. \\
Let $2N$ be the order of the system. The  homogeneous problem has N 
independent solutions verifying the boundary conditions at the center 
$\vec{Y}_1$, $\vec{Y}_2$, \ldots $\vec{Y}_N$  and  N independent solutions 
verifying the boundary conditions at the surface 
$\vec{Y}_{N+1}$, $\vec{Y}_{N+2}$, \ldots $\vec{Y}_{2N}$. 
  The corresponding fundamental matrix is
$M= (\vec{Y}_1, \vec{Y}_2,\ldots,\vec{Y}_N,\vec{Y}_{N+1},\ldots, \vec{Y}_{2N})$
 and the Green function matrix is given by:
\begin{eqnarray}
G_{ij}(r,r') &=& - \sum_{k=1}^N M_{ik}(r) M^{-1}_{kj}(r')  
\hspace{1cm}  r<r' \nonumber \\
&=&\sum_{k=N+1}^{2N} M_{ik}(r) M^{-1}_{kj}(r') \hspace{1cm}  r>r'
\end{eqnarray}
Equation (2) can also be written as
\begin{eqnarray}
\vec{Y}(r)&=&  -\sum_{k=1}^N \vec{Y}_k(r) \int_r^R \frac{|M_k(r')|}{|M(r')|} dr'
\nonumber \\ 
&+& \sum_{k=N+1}^{2N} \vec{Y}_k(r) \int_0^r \frac{|M_k(r')|}{|M(r')|} dr' \\
 &=& \sum_{k=1}^{k=2N} C_k(r) \vec{Y}_k(r) \nonumber 
\end{eqnarray}\
where $M_k$ is obtained by replacing the $k^{th}$ column of the fundamental 
matrix by $\vec{F}$. 

If the oscillations are excited by convection, the source term is different 
from zero below the observation level only and the coefficients of the 
solutions regular at the center are zero.

If the first component of $\vec{Y}$ is $\omega \delta r$ and the fourth one is 
$\delta L$, the power spectra for velocity and intensity fluctuations are 
given by
\[ |Y_i|^2= \sum_{j=N+1}^{2N}\sum_{k=N+1}^{2N}C_j(r)C_k^*(r)Y_{ij}(r)
Y_{ik}^*(r) \]
with $i$ equal to 1 and 4 respectively. Because of the summation, it is 
possible that the two spectra show different line asymmetries. If the Cowling 
approximation is used for non-radial oscillations, it is, for instance,
possible to choose $\vec{Y}_3$ such that $\delta r (R) = 0$ and $\vec{Y}_4$ 
such that $\delta L(R) = 0$. Then the line asymmetries for the velocity 
spectrum will be given mainly by the behaviour of $C_4$, while those of the 
intensity spectrum will be related mainly to the variation of $C_3$ with 
frequency.

If, as discussed above, the system is reduced to the second order, then eq. (4)
 reduces to $\vec{Y} = C_2(r) \vec{Y}_2$. There is no summation and if one  
spectrum is associated to each component of $\vec{Y}$, the two spectra 
 show indeed the same asymmetries. Notice that this result is independent 
of the explicit form of the  $A_{ij}$ and therefore of the hypothesis made to 
take dissipations into account, provided that a second order problem is 
obtained. 

Since it is necessary to solve the full non-adiabatic problem to explain the 
different line asymmetries of the two spectra, we can also wonder whether the 
 source depth obtained  from the velocity data only and using a second 
order problem can be trusted. 

After discussing the problem of velocity and intensity line asymmetries with a 
second order problem and failing to explain the opposite asymmetries indeed, 
Rast and Bogdan (1997a) have suggested that the problem can be solved if the  
two spectra have different noice levels. This might be the case but their 
remark mostly challenges observers who ought to remove properly the noice in 
their data analysis. If after this has been done and that informations 
concerning the excitation source have been clearly obtained, the two spectra 
still show opposite line asymmetries then the problem will have to be studied 
with the full fourth (in the radial case and in the nonradial one with the 
Cowling approximation) or sixth order system of non-adiabatic stability 
equations.

\section {The causes of line asymmetries}
We will analyze the causes of line asymmetries in the Cowling approximation 
since it is very good for 5 min. p-modes. Then the system (1) is of the fourth 
order and it is interesting to write eq. (4) as 
\begin{eqnarray}
\vec{y}(r) &=& -\vec{Y}_3(r)\int_0^r \frac{\vec{Y}_4(r').\vec{V}(r')}{|M(r')|}dr' \nonumber \\
 &+& \vec{Y}_4(r)\int_0^r \frac{\vec{Y}_3(r').\vec{V}(r')}{|M(r')|}dr'
\end{eqnarray}
with $\vec{V}$ given by:
\begin{eqnarray}
 && \{(T_{12}F_2-T_{13}F_3+T_{14}F_4),(-T_{21}F_1+T_{23}F_3-T_{24}F_4),
\nonumber \\
&&(T_{31}F_1-T_{32}F_2+T_{34}F_4), (-T_{41}F_1+T_{42}F_2-T_{43}F_3)\}  
\nonumber
\end{eqnarray}
$T_{ij}=T_{ji}$ is the determinant of the two by two matrix obtained from 
$(\vec{Y}_1,\vec{Y}_2)$ after suppressing lines $i$ and $j$. (we have kept the
four components of $\vec{F}$ though $F_3=0$ if the third line of eq. (1) is 
the transfer equation.) 

We will first assume that the excitation force is well localized and can be 
represented by a delta function and afterwards we will generalize the 
discussion to an extended source. 
 
First, we discuss the denominator $|M(r')|$. It cancels for each eigenvalues. 
Therefore, in the vicinity of one of them $\sigma= \sigma_R +i \sigma_I$, 
$|M(r')|= f(r',\sigma)(\omega-\sigma_R -i \sigma_I)$ with $f(r',\sigma) \neq 0$
 and for real $\omega$, $|M(r')|^2$ has minima close to the real part of the 
eigenvalues. Therefore it is nearly a periodic function (with a 
``period'' equal to the frequency separation between two successive line 
centers) but not exactly,  for two reasons:
\begin{enumerate}
\item the real parts of the eigenvalues are not exactly equidistant.
\item the extrema of $|M(r')|^2$ show a variation with frequency, especially 
close and above the cut-off frequency. This behaviour is already seen in simple
 idealized problems (Gabriel 1992).
\end{enumerate}
Therefore, close to the real part of an eigenvalue, $|M(r')|^2$ is symmetric 
but over a wider frequency range, it produces some skewness of 
the line profiles. However the asymmetries introduced by this term will be the 
same for the velocity and for the intensity spectra.

Let us now consider the numerator. \\
The two solutions regular at the surface vary slowly with frequency and 
they have generally been considered as constant in previous discussions. This 
is a good approximation for narrow lines. However, if line profiles are 
considered over a frequency range of the order of the eigenvalue separation, 
these two solutions will also introduce some skewness different for each 
spectrum. We now discuss the consequences of the variation of $\vec{V}$ 
assuming that $\vec{Y}_3$ and  $\vec{Y}_4$ are constant. Since $\vec{V}$ 
varies with the source type, the asymmetries will also be a function of the 
source type. 
Different spectra will have the same asymmetries only if $\vec{V}$ does not 
change its direction when $\omega$ varies, i.e. if  
$\vec{V}(r',\omega)= f(\omega)\vec{V_1}(r')$. This is very unlikely
 to occur as it requires very special properties for $\vec{Y}_1$, $\vec{Y}_2$
 and $\vec{F}$ and the velocity and intensity spectra will show different line 
asymmetries though not necessarily of opposite sign.
Because the radial and horizontal displacements of the eigenfunctions have 
one more node when the order is increased by one, the $T_{ij}$ show also an 
oscillatory behabiour with a quasi-period in frequency a little larger than 
twice the frequency separation between successive eigenvalues. However as 
these terms have to be squared to get the spectrum, they finally lead to a 
 function  with  a quasi-period in frequency a little larger than 
 the frequency separation between successive eigenvalues.
Again, because the real part of the eigenvalues are not exactly equidistant 
and the extremal values of the $T_{ij}$ vary with frequency, the numerator 
will produce some asymmetry in the lines if a large enough frequency domain is 
considered. But more important, the two quasi-periodic functions appearing in 
the numerator and denominator have different ``periods'' and they will be out 
of phase except, may be, for one or two lines. (The number of symmetric lines 
increases with the source depth and if it is close enough to the surface, there
 may be no such line.) This will be the main source of  asymmetry  
close to the center of the lines and the only one connected to the source 
position. Therefore to locate the source, it is better to study the  
asymmetries in a not too wide frequency domain around the line centers. Also, 
if a symmetric line is observed, the 
asymmetry presently under discussion must have opposite signs for lines on 
each side of the symmetric one.

In the case of an extended source, we must notice that the quasi-period of 
$\vec{V}$ changes with the position in the Sun; it increases as the point 
under consideration moves inward while that of the denominator does not change 
and therefore the phase lag between the numerator and the denominator changes 
too. Summing up the contributions of several layers 
will indeed smooth out the oscillations produced by the numerator and therefore
 reduce the amount of line asymmetries.

\section{Conclusions}
In this note, we have proved that the opposite line asymmetries observed in 
velocity and intensity power spectra cannot be explained by theories which 
approximate the dissipations in such a way to reduce the problem to a second 
order one. The solution of that problem requires the resolution of the full 
non-adiabatic problem. We have also analyzed the causes of line asymmetries in 
the frame of the general problem. The solution of that problem should be 
obtained including {\it all} the interactions between pulsation and convection 
(i.e. not just the perturbation of the convective luminosity) because they are 
all important in the upper layers of the convective enveloppe. There are 
several causes of asymmetry but 
only one of them can be related to the properties of the source. To pick up 
that one it is better to study line asymmetries not too far from line centers.

\end{document}